\documentclass[aip, reprint, amsfonts, amssymb, amsmath, superscriptaddress]{revtex4-2}
\usepackage[utf8]{inputenc}
\usepackage{ragged2e}
\usepackage{comment}
\usepackage{algorithm} 
\usepackage{algpseudocode} 
\usepackage{graphicx}
\usepackage{svg}

\newcommand{\rmd}{\ensuremath{\mathrm{d}}}

\newcommand{\super}[1]{\ensuremath{^{\mathrm{#1}}}}

\newcommand{\affilMSE}{\affiliation{Department of Materials Science and Engineering, Rensselaer Polytechnic Institute, Troy, NY 12180, USA}}
\newcommand{\affilMANE}{\affiliation{Department of Mechanical, Aerospace and Nuclear Engineering, Rensselaer Polytechnic Institute, Troy, NY 12180, USA}}
\newcommand{\affilECSE}{\affiliation{Department of Electrical, Computer and Systems Engineering, Rensselaer Polytechnic Institute, Troy, NY 12180, USA}}
\newcommand{\equalContrib}{\thanks{These authors contributed equally}}

\begin{document}

\title{First-principles molten salt phase diagrams through thermodynamic integration}

\author{Tanooj Shah}
\equalContrib
\affilMSE

\author{Kamron Fazel}
\equalContrib
\affilECSE

\author{Jie Lian}
\affilMANE

\author{Liping Huang}
\affilMSE

\author{Yunfeng Shi}
\affilMSE

\author{Ravishankar Sundararaman}
\email{sundar@rpi.edu}
\affilMSE

\begin{abstract}
Precise prediction of phase diagrams in molecular dynamics (MD) simulations is challenging due to the simultaneous need for long time and large length scales and accurate interatomic potentials.
We show that thermodynamic integration (TI) from low-cost force fields to neural network potentials (NNPs) trained using density-functional theory (DFT) enables rapid first-principles prediction of the solid-liquid phase boundary in the model salt NaCl.
We use this technique to compare the accuracy of several DFT exchange-correlation functionals for predicting the NaCl phase boundary, and find that the inclusion of dispersion interactions is critical to obtain good agreement with experiment.
Importantly, our approach introduces a method to predict solid-liquid phase boundaries for any material at an ab-initio level of accuracy, with the majority of the computational cost at the level of classical potentials. 
\end{abstract}

\maketitle

\section{Introduction}

Molten salts are a class of high-temperature ionic fluids that have recently attracted renewed interest due to their potential applications in modular nuclear reactors \cite{lebrunMoltenSaltsNuclear2007} and thermal energy-storage systems.\cite{zhangConcentratedSolarPower2013} 
A molten alkali halide salt such as LiF or a mixture such as FLiNaK may be used as a coolant instead of highly pressurized water in nuclear reactors; these salts can also act as the medium in which fuel and fission products are dissolved.\cite{grimesMoltenSaltReactorChemistry1970}

Accurate knowledge of the salt's phase diagram is critical for the design of such reactors. 
Experimental results serve as the ultimate benchmark of these properties, and methods such as CALPHAD can be used in the design process using parameters that are fit to experimental inputs.\cite{kroupaModellingPhaseDiagrams2013}
However such methods often use empirical functional forms for the relevant thermodynamic quantities needed to predict phase coexistence.
A more accurate method would obtain the relevant quantities - specifically, free energies - from a direct description of the interactions between the constituent atoms.

Predictions of thermodynamic properties of condensed phases from atomistic simulations can employ either Monte Carlo (MC)\cite{metropolisEquationStateCalculations1953} or molecular dynamics (MD)\cite{alderPhaseTransitionHard1957, verletComputerExperimentsClassical1967} approaches. 
Each method's accuracy depends upon the treatment of the interatomic potential energy function.
This potential energy function can be calculated from first principles using Kohn-Sham electronic density-functional theory (DFT) \cite{hohenbergInhomogeneousElectronGas1964, kohnSelfConsistentEquationsIncluding1965} in \emph{ab initio} molecular dynamics (AIMD) simulations, or approximated by classical force fields such as additive pairwise potentials.
MD simulations for predicting bulk phase equilibria accurately typically require system sizes containing at least $500 - 1000$ atoms, while AIMD simulations are typically limited by computational costs to $100 - 200$ atoms and time scales of $10 - 100$~ps.
Consequently, MD predictions of phase equilibria have typically employed classical force fields. \cite{eikeRobustGeneralMolecular2005, anwarCalculationMeltingPoint2003, meijerNovelProcedureDetermine1997}
However, this requires explicit parameterization of the empirical force fields for new materials, and even for single component systems, is limited in accuracy for smaller temperature and pressure ranges than a first-principles method.

Machine-learned interatomic potentials promise to bridge this gap between AIMD and classical MD by using highly flexible functional forms such as neural-network potentials (NNPs), which can better reproduce the potential energy surface from \emph{ab initio} results than simpler classical force-fields.\cite{behlerPerspectiveMachineLearning2016}
Several families of NNPs are finding increasing usage for MD simulations,\cite{lotPANNAPropertiesArtificial2020, leeSIMPLENNEfficientPackage2019, wangDeePMDkitDeepLearning2018} and essentially serve to extrapolate the DFT level predictions from smaller AIMD simulations to larger-scale MD simulations.
For molten salts, NNPs have been used to predict structure, diffusivity,\cite{leeComparativeStudiesStructural2021, liangMachineLearningDrivenSimulationsMicrostructure2021,liangTheoreticalPredictionLocal2021} shear viscosity,\cite{xuPowerfulPredictabilityFPMD2020} equations of state, heat capacity, thermal conductivity and phase coexistence at individual state points.\cite{liDevelopmentRobustNeuralnetwork2021}
However, to our knowledge, systematic mapping of the solid-liquid phase boundary of an alkali halide such as NaCl using either AIMD or machine-learned potentials in the pressure-temperature space has not been performed yet.

The most common way to estimate phase equilibria in MD is to carry out direct simulations of coexistence of the two phases in a large interface calculation. 
At a given state point, the interface will typically move to expand the thermodynamically favorable phase at the expense of the less favorable phase.\cite{frenkelUnderstandingMolecularSimulation2002}
This requires simulating large interfaces with at least $10^4$ atoms over long time scales (typically nanoseconds) and must be repeated over several state points to pinpoint the coexistence point. 
However at state points close to the true coexistence point, the velocity of the interface will typically be too low to reliably capture in simulations of tractable length.

A more accurate approach with better resolution involves calculating the free energy difference between the phases at various state points.
One can use thermodynamic integration \cite{kirkwoodStatisticalMechanicsFluid1935} (TI) in order to obtain these relevant free energies in molecular simulation. 
A reversible ``pseudosupercritical" pathway that directly transforms the liquid to the solid phase can be employed, where the interatomic potential $U(\lambda)$ is varied continuously as a function of an introduced path parameter $\lambda$, so as to establish a reversible transformation between the solid and liquid phases path at a particular state point $(P, T)$.
The resulting free energy difference between phases is calculated by integrating $\int \rmd\lambda (\partial U/\partial\lambda)$ along the pathway.
In particular, such a method avoids the problem of interfaces between two separate phases, and requires much smaller system sizes (on the order of 500-1000 atoms) than the aforementioned interface coexistence technique. 
TI simulations have recently been used with NNPs to assess solid-liquid coexistence for uranium, and solvation free energy predictions.\cite{kruglovPhaseDiagramUranium2019, jinnouchiMakingFreeenergyCalculations2020, fukushimaThermodynamicIntegrationNeural2019}
However to our knowledge no similar study has been yet performed using NNPs to compare the effects of different DFT approximations on an alkali halide phase boundary such as NaCl yet. 

Here, we introduce an approach using TI to combine low-cost classical force fields and more complex NNPs trained to electronic structure data in order to respectively combine the computational cost and accuracy advantages of each kind of interatomic potential.
Briefly, our method involves performing most of the complex transformations along the pseudosupercritical pathway using a cheap additive pairwise potential, and an additional bulk transformation from the classical potential to the NNP in each phase to obtain the NNP melting point.
From this initially determined melting point, we can extend the phase boundary in the (P, T) space using the Clausius-Clapeyron equation.

The remainder of this paper is organized as follows: in section~\ref{sec:Methods} we specify the classical interatomic potential and NNP parameterization details.
Afterwards, in section~\ref{sec:Approach}, we detail the phase equilibrium approach, starting from prediction of a single coexistence point using TI, and then using the Clausius-Clapeyron equation to extend the phase boundary in the (P, T) space.
Finally, we show results of our method in section~\ref{sec:Results} for the NaCl solid-liquid phase boundary, for NNPs trained to AIMD data with different choices of the exchange-correlation (XC) functional.
We find that the predicted phase boundaries are highly sensitive to the choice of XC functional, and that those functionals that explicitly build in treatment of dispersion interactions agrees with experiment over a much wider range of temperatures and pressures than other functionals.

\section{Methods}\label{sec:Methods}

\subsection{Fumi-Tosi Potential}

We perform classical MD simulations in LAMMPS \cite{plimptonFastParallelAlgorithms1995} using the standard Fumi-Tosi (FT) parameters \cite{fumiIonicSizesBorn1964} for NaCl. This model is also referred to as the rigid ion model (RIM) within the molten salts literature. Table~\ref{tab:fumi-tosi-params} lists the FT parameters used to model NaCl in this study.

\begin{equation} \label{eq:FT_function}
    U(r_{ij}) = Ae^{\frac{\sigma - r_{ij}}{\rho}} - \frac{C}{r_{ij}^6} + \frac{D}{r_{ij}^8} + \frac{k q_i q_j}{r_{ij}}
\end{equation}

In the present work, the long-range Coulomb part of the FT potential is treated using the damped shifted force model,\cite{fennellEwaldSummationStill2006} which allows faster computation than Ewald and particle-particle-particle-mesh (PPPM) methods.

While this simple functional form allows for rapid computation of interatomic forces, it limits the accuracy of predicted properties to specific chemical environments and thermodynamic conditions used to parameterize the model. \cite{luThermalTransportProperties2021}

\begin{table}[t] \centering
\caption{Fumi-Tosi parameters used for classical MD simulations of NaCl.\cite{fumiIonicSizesBorn1964}}
\label{tab:fumi-tosi-params}
\begin{tabular}{cccccc}
\hline\hline
Pair & $A$ (eV) & $\rho$ (\AA) & $\sigma$ (\AA) & $C$ (eV/$\AA^{6}$) & $D$ (eV/$\AA^{8}$) \\
\hline
Na-Na & 0.2637 & 0.317 & 2.340 &  1.0486 &   -0.4993 \\
Na-Cl & 0.2110 & 0.317 & 2.755 &  6.9906 &   -8.6758 \\
Cl-Cl & 0.1582 & 0.317 & 3.170 & 72.4022 & -145.4285 \\
\hline\hline
\end{tabular}
\end{table}

\subsection{Neural Network Potentials}

Neural networks are universal function approximators \cite{rumelhartLearningRepresentationsBackpropagating1986} that have been proven increasingly useful for describing the complex multidimensional potential energy surfaces of atomistic systems; in an NNP, the input to the neural network is a representation of the atomic coordinates, and the output is an energy.

\subsubsection{Featurization and neural network formulation}

All approaches to NNPs require some means to map the local neighbor configuration of each atom into the input features for the neural network.
It is particularly important for these features to account for rotational, translational and atomic permutation symmetries in order for the neural network to represent the potential energy landscape of the system with practical AIMD training data sets.
Prevalent current approaches for generating these features (or descriptors) from atomic configurations include
atom-centered symmetry functions,\cite{behlerAtomcenteredSymmetryFunctions2011}
smooth overlap of atomic positions (SOAP),\cite{bartokRepresentingChemicalEnvironments2013}
neighbor density bispectrum,\cite{bartokGaussianApproximationPotentials2010}
Coulomb matrices,\cite{ruppFastAccurateModeling2012}
and atomic cluster expansions (ACE),\cite{drautzAtomicClusterExpansion2019} amongst many others. 

Here, we use the SimpleNN code \cite{leeSIMPLENNEfficientPackage2019} for training and evaluating neural network potentials, which implements the atom-centered symmetry function approach.
In this approach, the total energy is written as a sum of atomic energies, each expressed as a neural network of several symmetry functions evaluated on the local atomic configuration.
These include radial functions that effectively measure the radial density of each atom type in a finite basis, and angular functions that similarly measure the angular distribution of pairs of atom types in a finite basis surrounding each atom.\cite{behlerAtomcenteredSymmetryFunctions2011}
We use the default set of radial $G_2$ and angular $G_4$ symmetry functions (70 total) implemented in the SimpleNN package with a cutoff of 6~\AA.

To train neural networks using the SimpleNN code, we use the built-in principal component preprocessing to mitigate linear dependence of symmetry functions and thereby accelerate the training.\cite{leeSIMPLENNEfficientPackage2019}
We also use adaptive sampling of the local atomic configurations based on Gaussian density functions, which increases the weight for infrequently encountered configurations in the loss function for training and improves the transferability of the resulting potential.\cite{jeongReliableTransferableMachine2018}
Finally, we find that a standard feed-forward neural network architecture with two hidden layers of 30 nodes each (30-30) proves sufficient, with negligible reduction of training errors with deeper or wider networks. 

\begin{table}
\caption{Structure, thermodynamic state points, and number of configurations sampled (10~fs apart) for each NaCl AIMD simulation used to train the NNPs utilized in this study.}
\label{tab:training-sets}
\begin{tabular}{lccc}
\hline\hline
Structure  & $T$ (K)  & $P$ (bar)  & \# configs \\
\hline
Rocksalt & 1000 & 1 & 201 \\
CsCl  & 1000 & 1 & 201 \\
Zincblende  & 1100 & 1 & 201 \\
Liquid & 1300 & 1 & 201 \\
Liquid & 1100 & 1 & 201 \\
Liquid & 1500 & 1 & 201 \\
Liquid  & 1700 & $10^{5}$ & 201 \\
Liquid  & 1500 & $5\times 10^{4}$ & 201 \\
Liquid & 1500 & $2\times 10^{4}$ to $3\times10^{6}$ & 108 \\
\hline\hline
\end{tabular}
\end{table}

\begin{figure}
\includegraphics[width=\columnwidth]{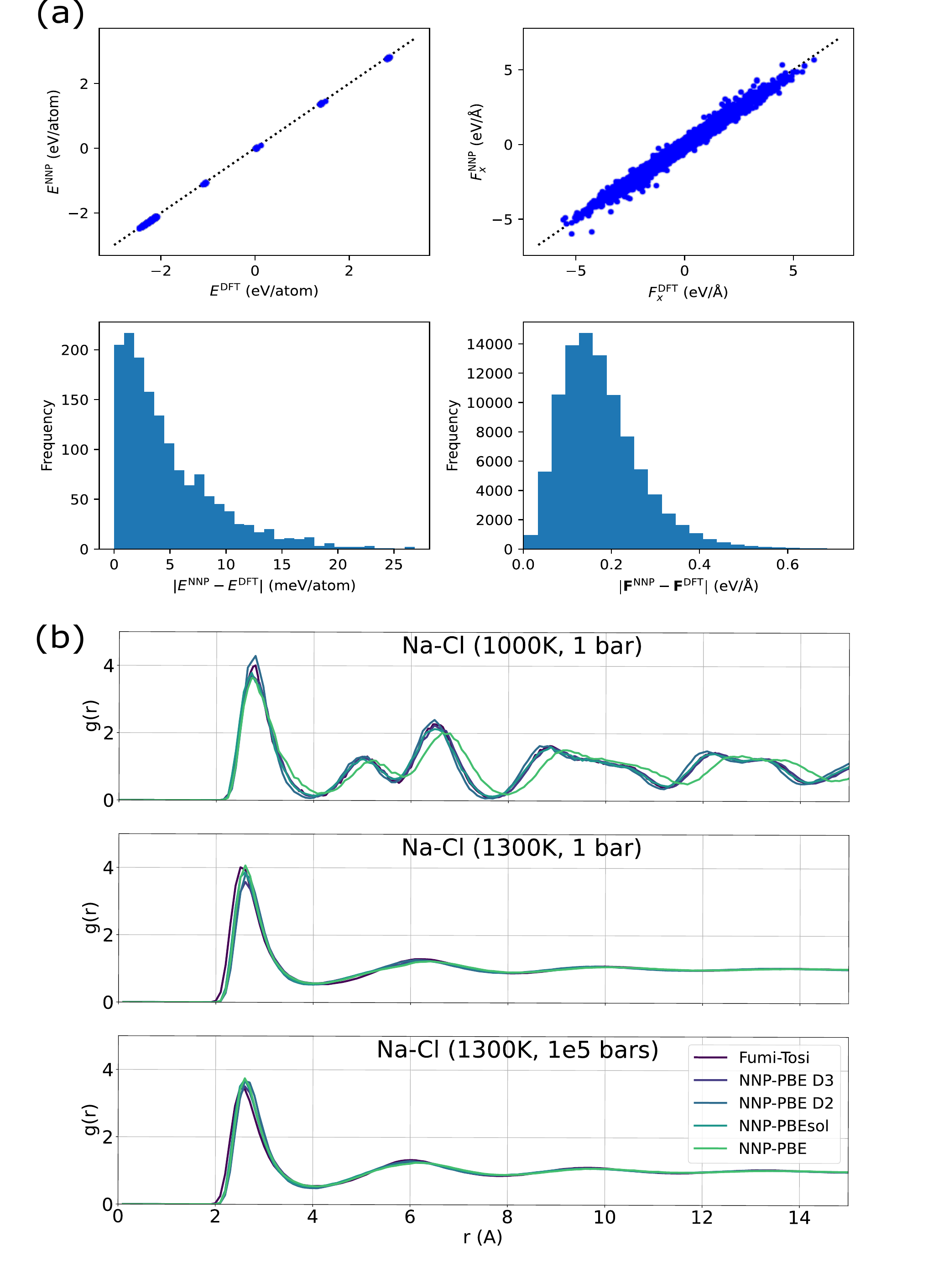}
\caption{(a) Correlation plots (above) and error distributions (below) between AIMD data and NNP for energies (left) and forces (right panels).
(b) Comparison of partial radial pair correlation functions between Fumi-Tosi and NNPs trained to AIMD with each of the four XC functionals considered here, for solid (top), liquid at ambient pressure (middle) and high pressure (bottom panels).
The predicted RDFs are overall very similar, except for overstructuring in the solid RDF from Fumi-Tosi potential compared to the NNPs, and slight shifts of the NNP-PBE RDF peaks to the right relative to the others due to its underbinding.
High pressure predominantly affects the liquid RDFs from the second coordination shell onwards.} 
\label{fig:nnp-performance}
\end{figure}

\subsubsection{AIMD Training Data}

An appropriate dataset of AIMD calculations is critical to reliably fit the parameters in the neural networks.
For the present use case, we need to ensure that the NNP is able to accurately model both solid and molten NaCl across a wide range of temperatures and pressures.
Table~\ref{tab:training-sets} lists the training AIMD simulations we use.
We include solid configurations in the stable rocksalt and metastable cesium chloride and zincblende structures near the melting temperature, and liquid configurations at ambient and high pressures spanning a range of temperatures above the melting point.
High pressure simulations are necessary in order to ensure stability of the NNPs, essentially by sampling more of the repulsive regime of the potential energy surface. \cite{liDevelopmentRobustNeuralnetwork2021}

In order to compare the effects of different functionals, we repeat the AIMD simulations, NNP training and all subsequent calculations for four different XC functionals, starting with the most frequently used Perdew-Burke-Ernzerhof (PBE) generalized-gradient approximation (GGA).\cite{perdewGeneralizedGradientApproximation1996}
Since PBE often underbinds solids leading to larger lattice constants, we also use its version reparameterized for solids, PBESol.\cite{perdewRestoringDensityGradientExpansion2008}
To analyze the impact of long-range dispersion interactions, we consider two variants of dispersion corrections to PBE, namely PBE D2\cite{grimmeSemiempiricalGGAtypeDensity2006} and PBE D3,\cite{grimmeConsistentAccurateInitio2010}
which have been shown to be important for accurate structure prediction in molten salts.\cite{royUnravelingLocalStructure2021}
For the remainder of this paper, we refer to these four NNPs trained to different XC functionals as NNP-PBE, NNP-PBEsol, NNP-PBE D2 and NNP-PBE D3 respectively.

Each AIMD simulation is started from an equilibrated configuration of 64 atoms using the Fumi-Tosi potential, with this size sufficient to get atomic environments extending to the symmetry function cutoff and to require no Brillouin zone sampling in the DFT.
The AIMD simulations are performed using the open-source JDFTx software,\cite{sundararamanJDFTxSoftwareJoint2017} using a Nose-Hoover thermostat and barostat, a time step of 1~fs, with configurations extracted for the data set every 10~fs. 
Configurations are extracted at this frequency in order for a balance between obtaining sufficient training data and tractable computation time for each individual simulation.  
Each AIMD simulation is run for 2 ps in the NPT ensemble, except for the high pressure sweep, which consists of four snapshots chosen along a classical MD compression simulation and then simulated in AIMD as an NVE ensemble for 0.2 ps each. 
We use a plane-wave basis with kinetic energy cutoffs of 20 and 100 Hartrees for wavefunctions and charge densities, respectively, as recommended for use with the GBRV ultrasoft pseudopotential set,\cite{garrityPseudopotentialsHighthroughputDFT2014} and converge the wavefunctions to an energy threshold of $10^{-7}$ Hartrees at each time step.
All subsequent simulations using the NNP are performed in LAMMPS. \cite{plimptonFastParallelAlgorithms1995}

\subsubsection{Benchmarks} \label{sec:Benchmarks}

We validate the trained NNP in two ways - by comparing its forces and energies to those generated via AIMD, and by comparing its radial distribution functions (RDFs) to those generated by the FT potential at three different state points.
This allows us to check if the forces and energies learned by the neural network are sufficient to capture the relevant structures of each phase needed for the later phase boundary calculations.  

Figure~\ref{fig:nnp-performance}(a) shows the correlation and errors between the NNP and AIMD (DFT) energies and forces.
The energy errors are all within 26 meV/atom, which is thermal energy $k_{B}T$ at ambient temperature.

The resulting structures predicted by the NNPs are checked by running larger calculations with each potential on 4096-atom simulation cells of the solid and liquid phases at different state points displayed in Figure \ref{fig:nnp-performance}(b).
In the rocksalt phase, NNP-PBE predicts a less dense phase than the other potentials (RDF peaks shifted slightly to the right), which is expected due to that functional's well-known underbinding in solids, however the RDFs for the other NNPs appear to overlap with FT.
Note that in the liquid phase, the RDFs for all four NNPs appear to overlap with FT at both ambient and high pressures, indicating that each potential has learned the appropriate structure.

\subsubsection{Cross-validation strategy} \label{sec:CrossValidation}

We also implement a 3-fold cross validation strategy to assess the impact of training errors of each NNP on the final phase boundary results presented later in this paper.
For each exchange-correlation functional, we fit three more NNPs to 2/3 of the AIMD training data, excluding a different 1/3 each time, and restart training with different randomly initialized weights in the network.
The different subsets are selected to evenly span the range of configurations in the training data.
A random sampling is not used to select the subsets of training data, in order that none of the subsets lose out any relevant information in the training data needed to maintain a stable potential. 

All subsequent calculations for the phase boundaries use four NNPs for each functional - the three NNPs using 2/3 of the training data, and one NNP using all the training data.
Subsequent estimates along the phase boundary are reported as the mean of the predictions made by each of these four NNPs, and error bars are reported as the standard deviation of these predictions.

\section{Phase Coexistence Approach} \label{sec:Approach}

\subsection{Direct Interface Coexistence}

The most commonly used technique for estimating solid-liquid coexistence in MD simulations is to directly set up an interface between a crystalline configuration and a liquid configuration and allow the system to equilibrate at a specified state point.\cite{allenComputerSimulationsLiquids2017}
If the state point is far from the phase boundary, the thermodynamically favorable phase grows at the expense of the less favorable one, whereas close to the phase boundary, the interface does not move appreciably.

\begin{figure}
\includegraphics[width=\columnwidth]{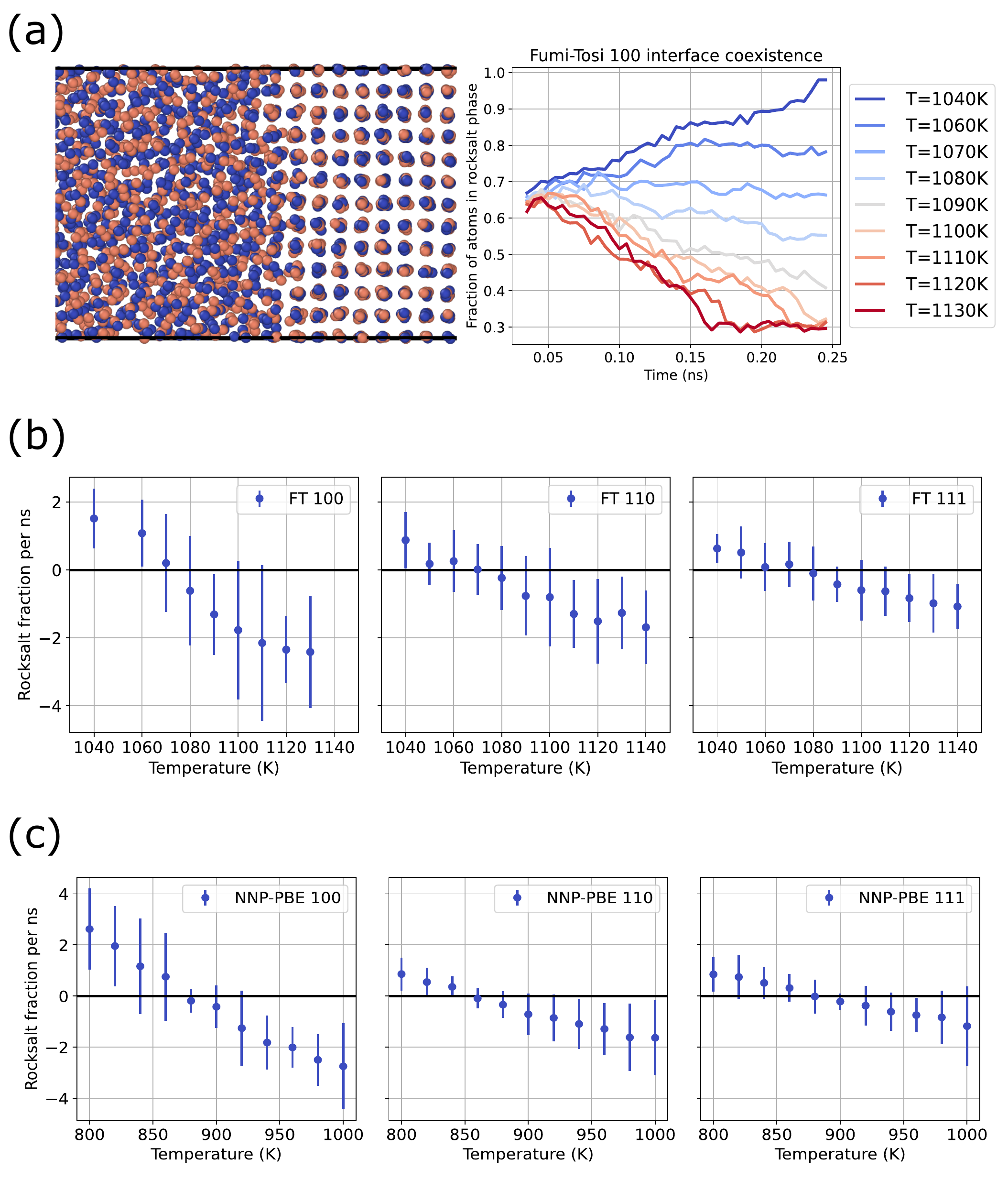}
\caption{(a) Left: a snapshot of a representative interface coexistence simulation, here with the interface set up along the (100) plane. Right: the fraction of atoms in the rocksalt phase as a function of simulation time for a simulation set up along the 100 interface using the Fumi-Tosi potential. (b) The rate of change of the fraction of atoms in the rocksalt phase as a function of temperature along the 100, 110 and 111 interfaces respectively, for the Fumi-Tosi potential. (c) The rate of change of the fraction of atoms in the rocksalt phase for the NNP-PBE potential. A coexistence temperature can be coarsely estimated from the zero-crossing of such curves, however the resulting estimate is inherently a feature of the surface set up in the simulation.}
\label{fig:coex-data}
\end{figure}

We test the direct interface coexistence method as an initial benchmark for the ambient-pressure melting temperature of the FT potential and the NNP trained to the PBE functional (hereafter referred to as NNP-PBE).
For each direct interface simulation, we start with a large orthorhombic supercell of NaCl on the order of $10^4$ atoms and convert the middle region to a liquid by running a high-temperature (3000~K) NVT simulation for 30~ps, while the remainder of the system is excluded from time-integration during this initial melt. 
We subsequently reset velocities of the entire system, and run NPT simulations to equilibrate the system at each candidate temperature at ambient pressure.
We monitor the fraction of atoms in the crystalline phase as a function of time with Steinhardt's $q_6$ order parameter\cite{steinhardtBondorientationalOrderLiquids1983}, and repeat this at several candidate melting temperatures. 
Figure \ref{fig:coex-data} a) displays a representative snapshot of the section of the system with the interface and the fraction of atoms in the solid phase as a function of time at different temperatures for the FT potential. 

An inherent drawback of this method is that any assessment of a coexisting state point depends upon the orientation of the interface set up between the solid phase and the liquid phase.
We repeat these simulations with solid surfaces oriented along the (100), (110) and (111) facets, and track the rate of change of solid fraction as a function of temperature for each, displayed in Figure \ref{fig:coex-data} b) and c) for the Fumi-Tosi and NNP-PBE potentials, respectively. 

While a rough value of $T_{m}$ can be estimated at 1070 $\pm$ 30 K and 875 $\pm$ 40 K for the FT and NNP-PBE potentials respectively, we see in Figure \ref{fig:coex-data} b) that there is a qualitative difference in behaviour for each facet, and hence that such a method may have limitations for estimating coexistence of the \emph{bulk} of a material with high accuracy. 
Moreover, the velocity of the interface is essentially a product of a mobility term and a driving force, and this driving force is proportional to $|T - T_{m}|$.
Close to the melting temperature, the velocity of the interface may be too slow to observe in a tractable simulation on the order of nanoseconds \cite{frenkelUnderstandingMolecularSimulation2002}.

\subsection{Free Energies Through Thermodynamic Integration}

Predicting phase coexistence from the free energy difference between phases as a function of thermodynamic variables is a more robust method, and can be achieved with significantly smaller systems than the ones necessary in direct interface coexistence simulations above.\cite{chipotFreeEnergyCalculations2007}

Most commonly, the free energy difference between two systems (or states) with different interaction potentials $U$ in molecular dynamics can be obtained via thermodynamic integration (TI),\cite{kirkwoodStatisticalMechanicsFluid1935} which involves the construction of a reversible pathway between the two states. 
In the canonical ensemble, we can obtain the Helmholtz free energy difference as 
\begin{equation}
\Delta A_{TI} = \int_0^1  d\lambda \left\langle \frac{\partial U}{\partial \lambda} \right\rangle. \label{eqn:TI}
\end{equation}
Here, $\lambda$ is a parameter that continuously changes the interaction potential from $U(\lambda=0)$ at the starting point to $U(\lambda=1)$ at the endpoint of the pathway, and the average $\langle\frac{\partial U}{\partial \lambda}\rangle$ is calculated in a canonical ensemble corresponding to each intermediate $\lambda$.
The key requirement is that the path is reversible, so that the ensemble averages are continuous with respect to $\lambda$.
Reversibility of the pathway is maintained by ensuring that the system is in equilibrium at each $\lambda$ point.

The resulting ensemble averages $\langle\frac{\partial U}{\partial \lambda}\rangle$ depend upon the way $\lambda$ is introduced to the interatomic potential energy function. 
Since the free energy is a state variable it does not matter whether the pathway or the intermediate states are physically realistic, and only the reversibility of the path is important; such simulations are often referred to as alchemical simulations in the community.

We can employ a ``pseudosupercritical'' pathway \cite{eikeRobustGeneralMolecular2005} to obtain the free energy difference between the solid and liquid phase at a single state point (P, T) for a given interatomic potential.
We can also use TI to transform from relatively expensive NNPs to cheaper additive pairwise potentials such as the FT potential at the start and end of said pathway. 
This allows us to obtain final results that only depend on the NNP, even though most of the computation over the pathway is performed with pair potentials, \emph{as long as the path remains reversible}.
Once an initial coexistence point is found, the phase boundary can be extended in the (P, T) space by integration of the Clausius-Clapeyron equation \cite{clausiusMotivePowerHeat1850} using data from simulations, for which numerous techniques are available in the literature.

So our overall scheme for predicting the first-principles solid-liquid phase boundary of NaCl using NNPs (trained to DFT) can be broken down into three major steps:
\begin{enumerate}
\item use a pseudosupercritical pathway at ambient pressure with the FT potential to calculate the solid-liquid free energy difference as a function of temperature $\Delta G_{sl} (T)$ to estimate that model's melting temperature $T_m\super{FT}$
\item use a thermodynamic cycle involving TI from NNP to FT to obtain $T_m\super{NNP}$
\item extend the phase boundary $T_m(P)$ by integrating the Clausius-Clapeyron equation.
\end{enumerate}
We detail each of the steps in the following sections, and note that the second and third steps are repeated for NNPs trained to different DFT functionals, and with different training sets for the cross-validation and error estimation strategy discussed in Section~\ref{sec:CrossValidation}.

\subsubsection{Pseudosupercritical Pathway for Fumi-Tosi $T_{m}$}

The melting temperature of NaCl at ambient pressure has been predicted previously for the FT potential using different TI approaches, including the pseudosupercritical pathway proposed by Eike et al, which yielded $T_m\super{FT} = (1089 \pm 8)$~K,\cite{eikeRobustGeneralMolecular2005}, and an approach proposed by Anwar et al, involving separate pathways connecting the solid to a harmonic crystal and the liquid to an ideal gas, which yielded $T_m\super{FT} = (1064 \pm 14)$~K.\cite{anwarCalculationMeltingPoint2003} The former approach originally computed the solid-liquid free energy difference $\Delta G_{sl}$ at a single guessed melting point, and used analytical corrections based on the solid and liquid equations of state to obtain a correction factor to predict the melting point.\cite{eikeRobustGeneralMolecular2005}

Since all of our subsequent more-expensive NNP steps depend on it, we adapt this pathway and make it more robust by running it independently at multiple temperatures to obtain $\Delta G_{sl} (T)$, and extract $T_m$ from its zero-crossing, which reduces the possibility of systematic errors in analytic corrections and convergence / ergodicity issues in individual calculations.

\begin{figure}
\centering
\includegraphics[width=\columnwidth]{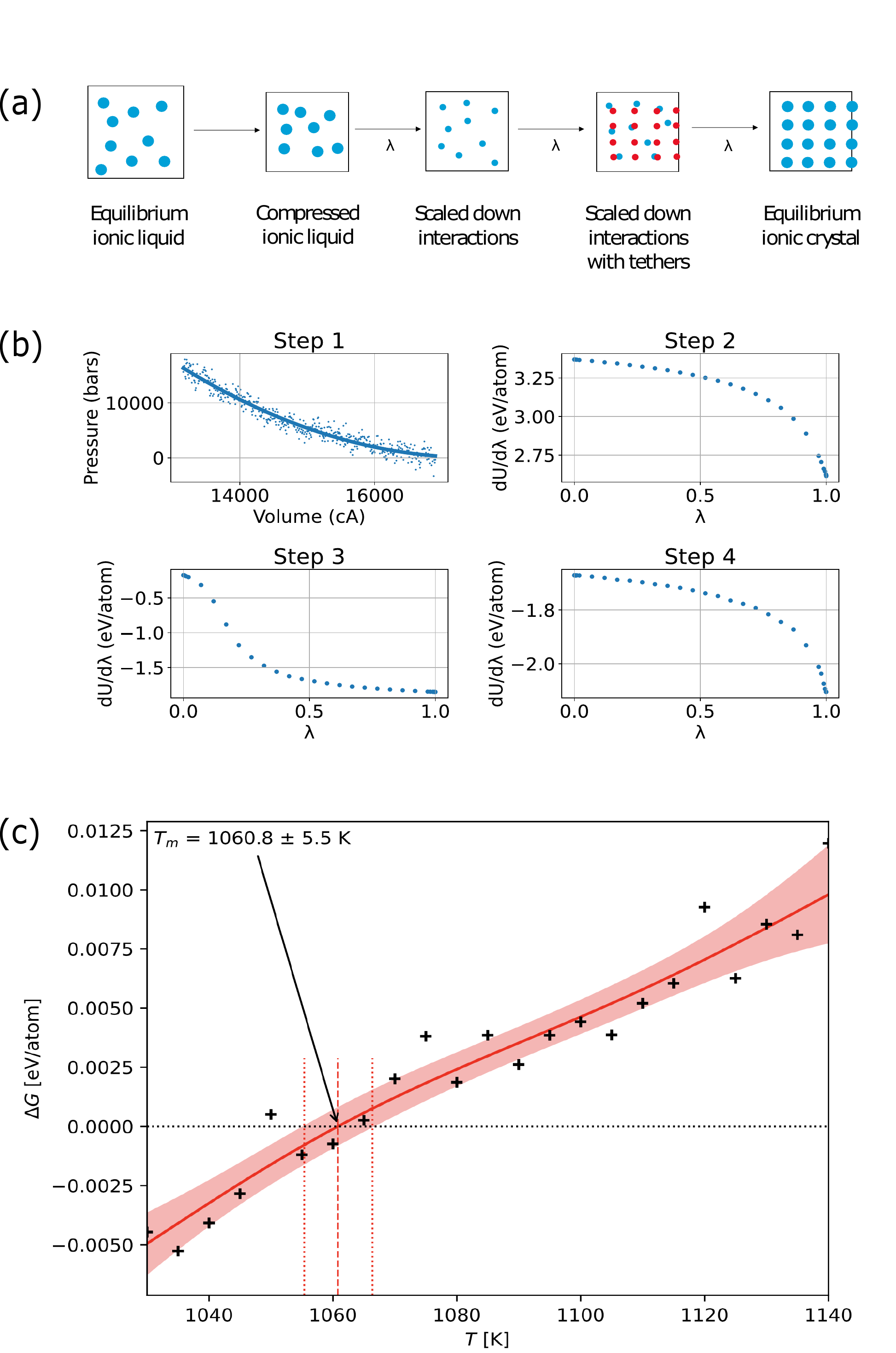}
\caption{(a) Thermodynamic integration pathway to compute solid-liquid free energy difference at a single state point. In the first step, an equilibrium ionic liquid is compressed to the corresponding equilibrium volume occupied by the crystal; in the second step, the ionic interactions are scaled down; in the third step, a tethering potential is switched on at each of the crystal lattice sites; in the fourth step, the tethering potential is switched off and the ionic interactions are scaled back to their full values.
(b) Representative Helmholtz free-energy curves for each of the four steps along the pathway, for a single run at $T = 1140$~K
(c) Gibbs free energy difference $\Delta G_{sl}(T)$ obtained from repeating this pathway for several temperatures.
To avoid assuming a specific polynomial form, we fit an ensemble of kernel ridge regression models to the data with resampling, and find the zero-crossing to get the melting point with a 95\% confidence interval, $T_m\super{FT} = 1060.8 \pm 5.5$~K.
}
\label{fig:classical-melt}
\end{figure}

The reversible pathway from liquid to solid at a single state point consists of 4 steps, displayed in Figure \ref{fig:classical-melt} a):
\begin{enumerate}
\item Deform liquid from its equilibrium volume to the equilibrium volume of solid at same (P, T), with free energy $\Delta A_{deform} = -\int_{V_{L}}^{V_{S}} P dV$.

\item Scale down Fumi-Tosi interaction potential $U_{FT}$ to $\eta U_{FT}$,
with free energy given by Eq.~\ref{eqn:TI} applied to $U(\lambda) = (1-\lambda) U_{FT} + \lambda (\eta U_{FT})$.
Using $\eta = 0.1$, this transforms the ionic liquid to a weakly interacting liquid, amenable for the next step of transformation to the solid's structure.

\item Switch on a tethering potential $U_{tether}$, consisting of attractive Gaussian potentials, $-A \exp(-B r^2)$, with $A = 2.0$~eV and $B = 1.1$~\AA$^{-2}$ at each crystal site, which interacts with the corresponding species of atoms.
This path has net potential, $U(\lambda) = (1-\lambda) (\eta U_{FT}) + \lambda (\eta U_{FT} + U_{tether})$,
with free energy given by Eq.~\ref{eqn:TI}, and transforms the weakly interacting liquid to an Einstein solid.

\item Restore the original Fumi-Tosi potential and simultaneously switch off the tethering potential.
This path has net potential, $U(\lambda) = (1-\lambda) (\eta U_{FT} + U_{tether}) + \lambda U_{FT}$,
with free energy given by Eq.~\ref{eqn:TI}, and transforms the Einstein solid to the ionic crystal.
\end{enumerate}

Adding the free energy from these four steps yields the net Helmholtz free energy difference between the solid and the liquid, $\Delta A_{sl}$.
We can then calculate the Gibbs free energy difference $\Delta G_{sl} = \Delta A_{sl} + P\Delta V_{sl}$, where $\Delta V_{sl}$ is the corresponding change in volume at this state point (P,  T).

We perform each of these steps in a cell with 256 Na-Cl ion pairs, in the $NVT$ ensemble using the Nose-Hoover thermostat in LAMMPS.
The simulations use a time step of 1~fs, and converged within 25~ps for each of 50 $\lambda$ values in steps 2 and 4, and within 50~ps for each $\lambda$ in step 3 above.
The final configuration at each $\lambda$ point is used as the initial configuration for the simulation at the next $\lambda$ point in order to ensure a smooth transformation along the pathway. 

We repeat this entire process to compute $\Delta G_{sl}(T)$ for several temperatures ranging from 1030~K to 1140~K, and fit an ensemble of kernel ridge models fit to different resamplings of the data to extract the zero-crossing with an error estimate, displayed in Figure \ref{fig:classical-melt}(b).
We thereby estimate the Fumi-Tosi melting point, $T_m\super{FT} = 1060.8 \pm 5.5$~K, which is consistent with the previous estimate from Ref.~\citenum{anwarCalculationMeltingPoint2003}, but is slightly lower than the one from Ref.~\citenum{eikeRobustGeneralMolecular2005}.
We use this ambient-pressure $T_m\super{FT}$ as starting point to determine the NNP melting temperatures.

\subsubsection{Thermodynamic Cycle to Obtain NNP $T_{m}$}

Once we have an estimate for the Fumi-Tosi melting point, we can use TI to obtain an estimate for the NNP melting point. 
We could adapt the approach mentioned in the previous section by simply adding on a bulk transformation between the NNP and the Fumi-Tosi potential at the start and end of the pseudosupercritical pathway (Figure \ref{fig:classical-melt} a) to obtain $\Delta G_{sl}\super{NNP} (T)$, however this would involve running multiple equilibrium simulations with the NNP at intermediate $\lambda$ points at every temperature - typical NNP simulations are on the order of 100 times slower than a FT simulation, so we propose an alternative approach which allows more rapid convergence for $T_m\super{NNP}$. 

\begin{figure}
\centering
\includegraphics[width=\columnwidth]{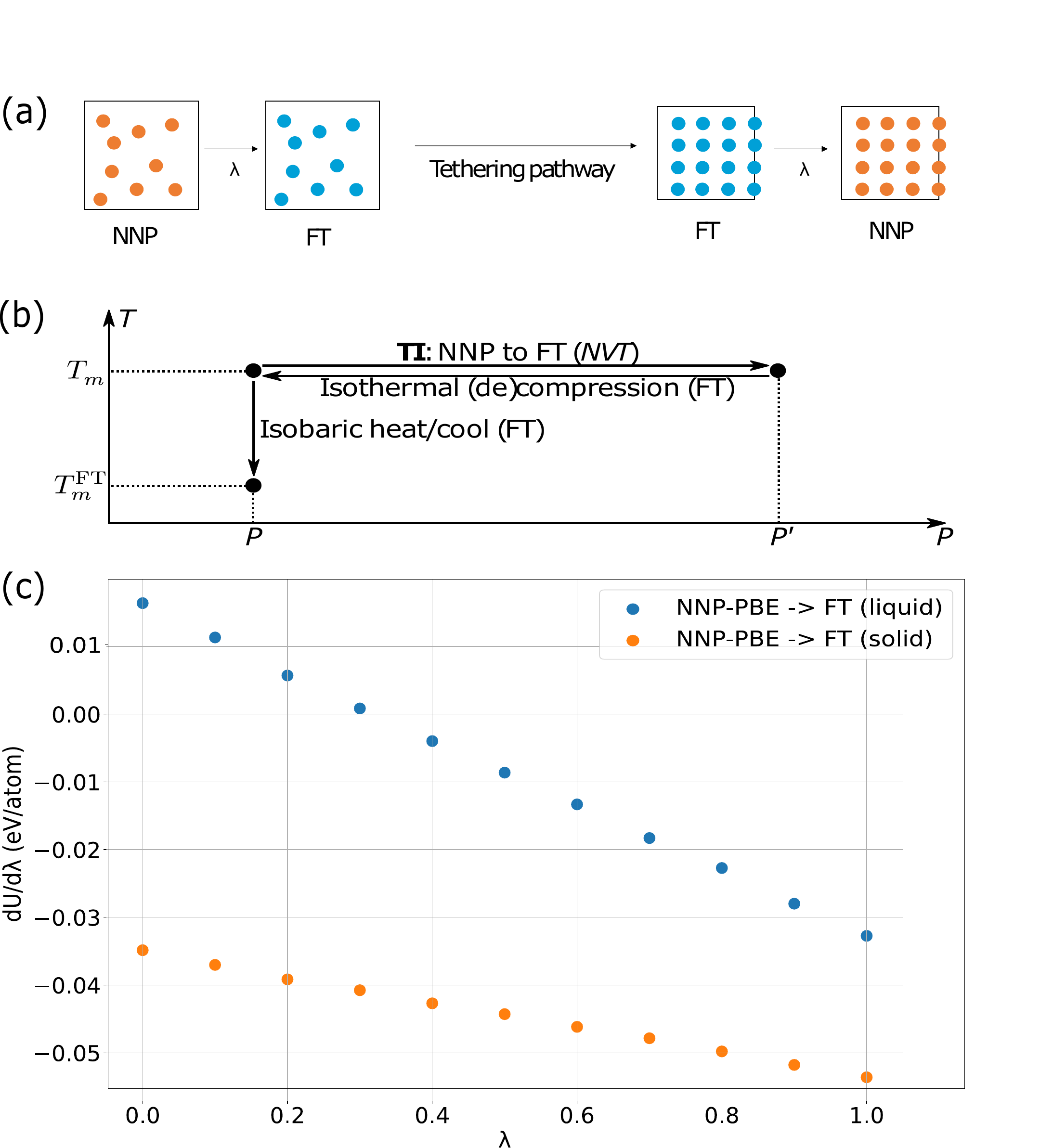}
\caption{
(a) The NNP melting point can in principle be found by performing bulk transformation to and from an additive pairwise potential at the endpoints of the pseudosupercritical pathway
(b) Thermodynamic cycle we use to iteratively converge upon $T_m\super{NNP}$. We make a guess for $T_m\super{NNP}$, use this cycle to compute $\Delta G_{sl}\super{NNP}$ at that state point, and subsequently update our guess using Equation \ref{eqn:DeltaTm}. This is a faster way to obtain $T_m\super{NNP}$ than the method in a) since it converges within a maximum of 5 iterations, whereas the pathway above to obtain $\Delta G_{sl}\super{NNP} (T)$ directly would involve running multiple expensive NNP simulations at various $\lambda$ points at every scanned temperature. 
(c) Representative $dU/d\lambda$ variation for the TI step above involving bulk transformation from the NNP to the the Fumi-Tosi potential; the nearly linear variation indicates that very few $\lambda$ points are needed to converge the free energy change.} 
\label{fig:FTtoNNP}
\end{figure}

Starting from an initial guessed value for $T_m\super{NNP}$, we run the following thermodynamic cycle \emph{separately} for each phase: 

\begin{itemize}
\item Convert NNP to Fumi-Tosi potential through TI, with $\Delta A$ computed using Eq.~\ref{eqn:TI} applied to $U(\lambda) = (1-\lambda) U_{NNP} + \lambda U_{FT}$. At fixed volume, this changes the pressure from $P$ for NNP equilibrium to $P'$ for FT equilibrium. Consequently, this step yields $\Delta G = \Delta A + V (P' - P)$.
\item Change pressure (at fixed $T$) of Fumi-Tosi solid/liquid from $P'$ back to $P$.
\item Change temperature (at fixed $P$) of Fumi-Tosi solid/liquid to $T_m\super{FT}$, where $\Delta G_{sl}\super{FT} = 0$ by definition 
\end{itemize}
See Figure \ref{fig:FTtoNNP} (b) for an elucidation of this cycle. 
Adding these three steps together, we obtain $\Delta G_{sl}$ for the NNP at the guessed temperature.
We can subsequently calculate a correction factor to update the guess for $T_m\super{NNP}$ using the following equation. 

\begin{equation}
\Delta T_m = -\frac{\Delta G_{sl}\super{NNP}}{\partial\Delta G_{sl}\super{NNP}/\partial T_m}
\approx T_m \frac{\Delta G_{sl}\super{NNP}}{\Delta H_{sl}\super{NNP}},
\label{eqn:DeltaTm}
\end{equation}
because $\partial G/\partial T = -S$ and $\Delta S_{sl} = \Delta H_{sl}/T_m$ at the true melting point, and the latter is approximately true close to the melting point.
We find that for the NNPs trained in this study, this pathway converges to 4~K tolerance within at most 5 such steps.

Figure~\ref{fig:FTtoNNP}(c) shows a representative $dU/d\lambda$ curve obtained for the NNP to FT connection with ten $\lambda$ points.
Note the near perfect linearity of $dU/d\lambda$ with $\lambda$, indicating that this TI can be performed with very few $\lambda$ points, possibly even with just three $\lambda$ points at 0, 0.5 and 1.
Once again, this indicates the power of the present approach to keep most of the computation at the cheaper classical potential level, requiring very few calculations using NNPs.
The main requirement is that the classical potential is just accurate enough to predict a liquid and solid phases that remain stable through the TI paths shown above.

\subsubsection{Extension of Phase Boundary using Clausius-Clapeyron Equation}

Once we have an initial point of solid-liquid coexistence $(P, T)$, we can numerically integrate the Clausius-Clapeyron equation,\cite{clausiusMotivePowerHeat1850}
\begin{equation}
\frac{dP}{dT} = \frac{\Delta H_{sl}}{T \Delta V_{sl}},
\label{eqn:ClausiusClayperon}
\end{equation}
to find the entire coexistence line in $P$-$T$ space for whichever interatomic potential.
Here, the solid-liquid difference in enthalpy $\Delta H_{sl}$ and molar volume $\Delta V_{sl}$ can be obtained directly from $NPT$ simulations of both phases at a known coexisting state point $(P, T)$.
This allows an initial estimate for the melting temperature at pressure $P' = P + \Delta P$ at $T' = T + \Delta P / (dP/dT)$ (we use a $\Delta P$ of 1000 bars in the present work). 

We then converge from this straight-line approximation to find the point where $\Delta G_{sl} = 0$, using a method very similar to the calculation of the NNP melting point in the previous section.
We run a compression step from $P \rightarrow P'$ at constant $T$, and then iteratively run heating steps from $T \rightarrow T'$ at constant $P'$ until convergence, using Equation~\ref{eqn:DeltaTm} at each iteration to obtain the correction factors. 
Note that these are essentially the last two steps of the thermodynamic cycle in Figure~\ref{fig:FTtoNNP}(b).

We find that this approach converges within 3 iterations for each step in pressure, with a convergence criterion of $|\Delta T| < 4~K$, for all the interaction potentials used in this study.
This approach is closely related to the coexistence-line free-energy difference integration method,\cite{meijerNovelProcedureDetermine1997} but distinct from Gibbs-Duhem integration;\cite{kofkeGibbsDuhemIntegrationNew1992} the present approach keeps each molecular dynamics simulation as an $NVT$ or $NPT$ simulation at a single state point for robustness and ease of applicability to both classical potentials and NNPs.

\section{Phase boundary results} \label{sec:Results}

\begin{figure}
\includegraphics[width=\columnwidth]{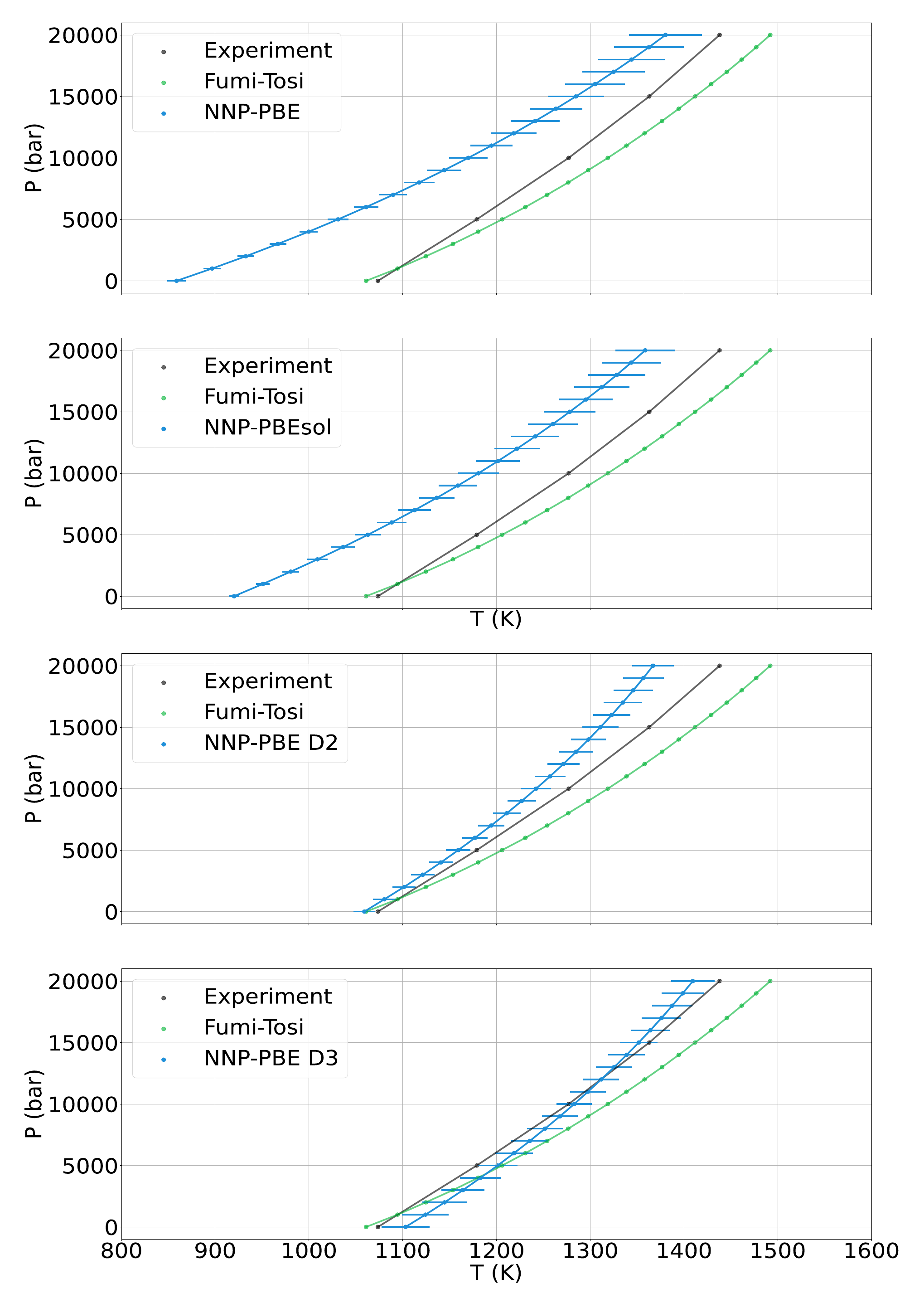}
\caption{Predicted NaCl solid-liquid phase boundaries for NNPs trained to four different DFT functionals, compared to experimental results and Fumi-Tosi predictions.
Error bars shown for the NNP predictions are from cross-validation using NNPs trained to different subsets of the DFT data for each case.
The NNPs trained to PBE and PBEsol without explicit dispersion corrections strongly underestimate the melting temperature at all pressures, while the PBE D2 and PBE D3 dispersion-corrected results agree better with experiment than the empirical Fumi-Tosi potential.}
\label{fig:PhaseBoundary}
\end{figure}

The techniques developed in the previous section allow mapping of the $P$-$T$ solid-liquid coexistence curve for any interatomic potential, including classical potentials such as the Fumi-Tosi potential for NaCl and machine-learned potentials, including NNPs.
Figure~\ref{fig:PhaseBoundary} compares the NaCl phase boundaries predicted by NNPs trained to four different DFT XC functionals against experimental measurements and the Fumi-Tosi classical potential predictions.

First note that while the Fumi-Tosi potential is accurate for the melting point at ambient pressure compared to experiment, it deviates from experiment at higher pressures, consistent with previous classical potential simulations.\cite{eikeRobustGeneralMolecular2005}
Note that even the slope $dP/dT$ of the coexistence line is incorrect near ambient pressure, indicating that the error stems from either the predicted enthalpy difference or molar volume difference between the phases, as indicated by the Clausius-Clapeyron equation.
Table~\ref{tab:crystal-benchmarks} shows that the Fumi-Tosi potential is reasonably accurate for the enthalpy difference and solid volume, but overestimates the liquid volume and thereby results in a smaller $dP/dT$ than experiment.

The NNP-PBE potential leads to a consistently lower melting point than experiment for all pressures, as seen in Figure~\ref{fig:PhaseBoundary}(a).
This is expected given the tendency of PBE to underestimate binding in solids generally.
Specifically, for NaCl, PBE predicts an almost 2\% larger lattice constant for the crystal than experiment, and understimates the atomization energy by 6\% (Table~\ref{tab:crystal-benchmarks}), leading to the $\sim 15$\% underestimation of the melting point.

The PBEsol functional is a reparameterization of PBE which restores the correct gradient expansion of the correlation energy, generally improving performance for solids.\cite{perdewRestoringDensityGradientExpansion2008}
This fixes the lattice constant of the crystal ($<0.2$\% error), but the atomization energy is still underestimated by 3\%.
Correspondingly, the NNP-PBEsol melting point predictions shown in Figure~\ref{fig:PhaseBoundary}(b) are slightly improved compared to the PBE case, but are still substantially lower than experiment at all pressures .

Only NNPs trained to DFT that includes dispersion corrections predict melting points that agree reasonably with experiment, shown in Figures~\ref{fig:PhaseBoundary}(c) and (d). 
This was also recently pointed out for the ambient-pressure melting point in a previous work\cite{liDevelopmentRobustNeuralnetwork2021}.
The dispersion-corrected PBE D2 variant\cite{grimmeSemiempiricalGGAtypeDensity2006} has the lowest error for this ambient-pressure melting point, but the PBE D3 variant\cite{grimmeConsistentAccurateInitio2010} exhibits better accuracy overall for the entire range of pressures considered in this study.

Table~\ref{tab:crystal-benchmarks} indicates that both PBE D2 and PBE D3 are actually less accurate than PBEsol for the lattice constant at lower temperatures; they are only more accurate for the atomization energy.
However both these functionals are more accurate for the solid and liquid volumes near the melting point, so the relative underbinding of these dispersion-corrected functionals for the perfect crystal become less important for solids at higher temperatures and for the liquid phase.
PBE D3 has both the closest molar volumes and enthalpy difference compared to experiment amongst all the functionals considered here (including the Fumi-Tosi potential), correlating with its best accuracy for the phase boundary across the pressure-temperature space.

\begin{table}[t]
\caption{Comparison of lattice constant $a$ and atomization energy $E_a$ of NaCl crystals at ambient temperature, as well as molar volumes and solid-liquid enthalpy difference at the respective melting points, predicted by different DFT exchange-correlation functionals and classical potentials against the experimental values.\cite{grayAmericanInstitutePhysics1972, brownLevelBornHaber2000} Values for $dP/dT$ are reported at ambient pressure.} 
\label{tab:crystal-benchmarks}
\begin{tabular}{ccc | ccccc}
\hline\hline
& $a$ & $E_a$ & $V_s$ & $V_l$ & $\Delta H_{sl}$ & $dP/dT$ \\
& (\AA) & (eV) & (L/mol) & (L/mol) & (kJ/mol) & (bar/K) \\
\hline
Experiment & 5.60 & 6.68 & 32.0\cite{kirshenbaumDensityLiquidNaCl1962} & 37.6\cite{kirshenbaumDensityLiquidNaCl1962} & 28.0\cite{ullmannUllmannEncyclopediaIndustrial1985} & 4689 \\
Fumi-Tosi  & 5.62 & 6.53* & 31.8 & 41.2 & 28.5 & 3043 \\
PBE        & 5.70 & 6.28 & 32.8 & 44.4 & 28.5 & 2438 \\
PBEsol     & 5.61 & 6.47 & 30.2 & 41.2 & 28.2 & 2550 \\
PBE D2     & 5.66 & 6.75 & 30.3 & 36.9 & 33.0  & 4944 \\
PBE D3     & 5.66 & 6.66 & 31.7 & 37.9 & 28.8 & 4653  \\
\hline\hline
\end{tabular}
\begin{flushleft}
*{\footnotesize Predicted energy of splitting crystal to ions,
combined with experimental Na ionization energy and Cl electron affinity,
since this classical potential can only describe ions and not atoms.}
\end{flushleft}
\end{table}

\section{Conclusion}

We introduced a computational approach to efficiently predict \emph{ab-initio}-level solid-liquid phase boundaries in molten salts using a combination of machine-learned potentials and thermodynamic integration.
We used NNPs trained to DFT with different exchange-correlation functionals in order to compare the accuracy of different DFT methods for the thermodynamics of molten salts, with error bars on all predictions using ensembles of NNPs trained to different \emph{ab initio} MD data.

Most importantly, we tailored the thermodynamic integration approach to carry out most of the simulations using low-cost classical potentials, with NNPs used only in the final connection.
Critically, once this approach is converged, the final result depends only on the NNP interaction potential, even though we used the lower-level classical potential in all intermediate steps of the path connecting the solid and liquid.
Overall, this approach makes it much more tractable to explore molten salt equilibria with accuracy ultimately limited by the first-principles methods underlying the NNPs.

Specifically, for the melting of NaCl, we show that treatment of long-range dispersion interactions in the DFT exchange-correlation functional is critical, with PBE D3 yielding the overall highest accuracy for solid-liquid coexistence across a wide range of pressures.
We show that the atomization energy of the crystal is the best proxy for the accuracy of melting-point predictions, while estimates of under/overbinding based on lattice constants do not correlate as well: PBEsol yields the best lattice constant, but significantly underestimates melting points at all pressures.
The overall approach described here was prototyped using NaCl as a model system, but is applicable for any single-component system with an interatomic potential that is accurate enough to exhibit a stable solid and liquid phase in the relevant temperature range.
Importantly, the thermodynamic integration approach removes dependence of the final results on this potential and allows prediction via NNPs using any underlying first-principles method.
Future work can extend such free-energy methods combining NNPs and thermodynamic integration to predict phase diagrams of binary systems and solubility limits from first principles.

\section*{Acknowledgements}
This work was supported by funding from the DOE Office of Nuclear Energy's Nuclear Energy University (NEUP) Program under Award \# DE-NE0008946.

\section*{References}
\bibliographystyle{apsrev4-2}
\bibliography{MoltenSalt}
\end{document}


\title{Supplementary information for:\\
`First principles molten salt phase diagrams through thermodynamic integration'}
\author{Tanooj Shah}
\equalContrib
\affilMSE

\author{Kamron Fazel}
\equalContrib
\affilECSE

\author{Jie Lian}
\affilMANE

\author{Liping Huang}
\affilMSE

\author{Yunfeng Shi}
\affilMSE

\author{Ravishankar Sundararaman}
\email{sundar@rpi.edu}
\affilMSE
\maketitle

\section*{Included scripts}

The accompanying archive of scripts contains code to reproduce the relevant free energy calculations as implemented in LAMMPS, including
\begin{itemize}
\item the reversible pathway to continuously transform between a solid and liquid using a pairwise potential to predict the classical potential's melting point
\item the self consistent series of simulations including bulk transformation between the NNP and pairwise potential to obtain the NNP's melting point at ambient pressure
\item the predictor-corrector algorithm to numerically integrate the Clausius-Clapeyron equation for extension of the phase boundary in the pressure-temperature space
\end{itemize}

This code can also be accessed at \filename{github.com/tanooj-s/nnpPhaseBoundary}. 

Each of the following sections present computational details referring to the path of the code and/or data files within the archive, relative to the directory for each step (named \filename{EikePathway/}, \filename{NNPMeltingPoint/} and \filename{ClausiusClapeyron/} respectively).

Within each directory above is also included four neural network potential files each trained to a different exchange correlation functional; these files are named \filename{PBE}, \filename{PBESol}, \filename{PBED2} and \filename{PBED3}. These potentials were generated using the SimpleNN software package, and the LAMMPS executable running the molecular dynamics simulations must be compiled to support the \filename{pair\string_style nn} command consistent with this flavor of NNP. If desired, the appropriate \filename{pair\string_style} and \filename{pair\string_coeff} lines in each \filename{.in} file can be edited to use different sorts of NNP packages such as DeePMD, so long as they are compatible with \filename{pair\string_style hybrid/scaled}. 

\section*{Reversible pathway between solid and liquid to obtain $\Delta G_{sl}(T)$}

Within the \filename{EikePathway/} directory, the \filename{masterScript.py} is run to perform simulations along the alchemical pathway that directly transforms the liquid to the solid at each temperature. This Python script invokes various Slurm job files (with the suffix \filename{.job}) necessary along each stage of the pathway. Each job file calls a corresponding LAMMPS input script with the same name (with the suffix \filename{.in}). Note that only the \filename{.in} scripts are included in this archive; job scripts will need to be customized for whatever cluster these simulations are being run on. Also note that due to the nature of this particular pathway the TI simulations cannot be run in parallel and must be run sequentially. A \filename{parseLammpsLog.py} script is also included here which converts each LAMMPS output log file into a \filename{.csv} file, which allows for automation of relevant calculations within \filename{masterScript.py} in between calling the various simulation scripts at different stages. The relevant LAMMPS scripts in this directory are as follows:
\begin{itemize}
\item \filename{create\string_solid.in} and \filename{create\string_liquid.in}, which create equilibrated ambient pressure solid and liquid Fumi-Tosi NaCl configurations at each temperature.
\item \filename{deform\string_liquid.in}, the first step of this pathway, which reads in the equilibrium liquid configuration and deforms it to the corresponding solid's volume.
\item \filename{ft2short.in}, the second step of this pathway, which scales down the ionic interactions by a factor $\eta$. Note that this script needs to be invoked sequentially from within \filename{masterScript.py} to be run at various $\lambda$ points, and that the starting configuration at each $\lambda$ is the datafile written out from the previous point. 
\item \filename{rocksalt\string_lattice.in}, which runs a single step simulation to generate lattice sites corresponding to the equilibrated solid volume at each temperature, and \filename{addWellAtoms.py} which writes a data file with these generated lattice sites along with the final liquid configuration generated by the previous step (i.e. the data file generated by \filename{ft2short.in} at $\lambda=1$); these create the tethering sites necessary for the next two steps of this pathway.
\item \filename{tetherOn.in}, the third step of this pathway, which switches on the tethering potential at the lattice sites. Note that like \filename{ft2short.in} this must be run sequentially at each $\lambda$ point for a smooth transformation. 
\item \filename{fumiOnTetherOff.in}, the fourth step of this pathway, which switches off the tethering potential and scales back up the ionic interactions to their full values. Like the previous two steps along the pathway, this is run sequentially at different $\lambda$ points with initial configurations at each point using those generated by the previous $\lambda$ point. 
\end{itemize}

When this pathway is finished running along the selected temperatures, a \filename{delA-T.csv} file is generated which contains various free energies as a function of temperature. The \filename{dGpath} column can be plotted against the \filename{Trun} variable to obtain the relevant $\Delta G_{sl}(T)$ curve, from which the melting point can be obtained by finding the zero-crossing of the function. 

An \filename{lmp\string_srl} variable is defined at the beginning of \filename{masterScript.py}, which should be edited to point to the relevant LAMMPS executable these simulations are being run with. \filename{time.sleep()} is used liberally in this and the following Python scripts to pause execution while relevant simulations are being run.

\section*{Prediction of an NNP melting point}
Similar to the previous directory, within the \filename{NNPMeltingPoint/} directory a \filename{masterScript.py} file launches all necessary Slurm jobs and performs the requisite intermittent calculations. Four flags must be passed into this script which are as follows: \filename{T0}, which is the classical potential's melting temperature at ambient pressure obtained from the previous pathway; \filename{Tg}, which is the initial guess for the NNP melting temperature at the same pressure; \filename{pot}, which is the location of the neural network potential file; \filename{t}, which is the temperature convergence threshold; and \filename{s}, the supercell size of the system to use. For NaCl, which has 8 atoms in a unit cell, \filename{s=4} leads to systems of 512 atoms and \filename{s=5} has 1000 atoms; we recommend using systems with at least 500 atoms for appropriate bulk properties. We used a threshold of 4K which led to covergence of the NNP's predicted melting point within 4 or 5 iterations for each trained NNP.

As above, the \filename{.job} files are not included in this directory. The relevant LAMMPS \filename{.in} files are as follows:

\begin{itemize}

\item \filename{create\string_fumi\string_solid.in} and \filename{create\string_fumi\string_liquid.in}, which create ambient pressure Fumi-Tosi NaCl configurations at that potential's melting point

\item \filename{create\string_nn\string_solid.in} and \filename{create\string_nn\string_liquid.in}, which create ambient pressure NNP configurations at each iteration of the guessed melting point

\item \filename{ti\string_solid.in} and \filename{ti\string_liquid.in} which perform bulk transformation from the NNP to the Fumi-Tosi potential separately for each phase and store the log files at different values of $\lambda$ within the subdirectories \filename{solid/} and \filename{liquid/} respectively

\item \filename{solid\string_press.in} and \filename{liquid\string_press.in}, which take the Fumi-Tosi NaCl configurations generated at the end of the TI step back to ambient pressure $P_{0}$

\item \filename{solid\string_heat.in} and \filename{liquid\string_heat.in}, which take the Fumi-Tosi NaCl configurations generated at the end of the previous step back to the known melting temperature $T_{0}$, where $\Delta G_{sl}=0$ by definition

\end{itemize}

At a single guessed NNP melting point temperature $T_{g}$, these simulations yield a $\Delta G_{sl}$ along this pathway, which is used along with ${\Delta H}$ between the Fumi-Tosi configurations at $T_{0}$ (that is, the latent heat of melting for that potential) to obtain a correction factor $\Delta T$ used to update $T_{g}$, which as mentioned above we found to converge within 4 or 5 iterations. An \filename{nnp-predictions.csv} file is generated with each iteration of the predicted NNP melting temperature as the rows; the final value is the converged melting point.

\begin{algorithm}[H]
\caption{Method to predict NNP $T_m$.\label{algorithm:NNP_TI}}
\begin{algorithmic}[1]
\State With known Fumi-Tosi melting point $(P^{0},T^{0})$
\State Set temperature convergence threshold 
\State Make an initial guess for the NNP melting temperature at ambient pressure $(P^{0},T_{g})$
\State Set initial high $\Delta T$
\While 
{$\Delta T > threshold$}
    \State Run NVT TI simulations $(P^{0},T_{g})_{NNP} \rightarrow (P^{\lambda},T_{g})_{FT}$, calculate $ \Delta G_{TI}$
    \State Run Fumi-Tosi (de)compression $(P^{\lambda},T_{g}) \rightarrow (P^{0},T_{g})$, calculate $\Delta G_{C} $
    \State Run Fumi-Tosi heating (cooling) $(P^{0},T_{g}) \rightarrow (P^{0},T^{0})$, calculate $\Delta G_{H} $
    \State Use $ \Delta G_{TI} + \Delta G_{C} + \Delta G_{H}$ to obtain correction factor $\Delta T$
    \State Refine guess for NNP melting temperature $T_{g} = T_{g} + \Delta T$
\EndWhile
\State Predict converged NNP melting temperature $T^{m}_{NNP} = T_{g}$ at $P^{0}$
\end{algorithmic}
\end{algorithm}

\section*{Phase boundary extension in (P, T)}

Within the \filename{ClausiusClapeyron/} directory, the \filename{masterScript.py} can be run to extend the phase boundary of either an NNP or the pairwise Fumi-Tosi potential. The flags to pass into this script are: \filename{T0}, the predicted melting temperature for whichever potential; \filename{P0}, the coexistence pressure corresponding to that temperature; \filename{dP}, the interval of pressures at which to predict melting temperatures at; \filename{c}, the number of melting points to predict; \filename{s}, the system supercell size (which should be the same as in the previous step); \filename{pot}, the location of the NNP potential file (if predicting the Fumi-Tosi potential this flag is moot, however the relevant \filename{pair\string_style} and \filename{pair\string_coeff} lines in the subsequent scripts should be edited appropriately); and \filename{t}, the temperature convergence threshold. 
The relevant LAMMPS \filename{.in} files are as follows:

\begin{itemize}
\item \filename{create\string_solid.in} and \filename{create\string_liquid.in}, which create solid and liquid configurations at each converged coexisting state point
\item \filename{solid\string_compress.in} and \filename{liquid\string_compress.in}, which run compression simulations from each converged melting pressure to the next pressure the melting temperature is to be predicted at
\item \filename{solid\string_heatup.in} and \filename{liquid\string_heatup.in}, which run heating simulations from each converged melting temperature to the predicted melting temperature at the next state point

\end{itemize}

\filename{masterScript.py} iteratively runs the heating simulations until the correction factor $\Delta T$ derived from $\Delta G_{sl}$ along this pathway converges within the given threshold; it then creates solid and liquid configurations at that converged state point, runs the compression simulations to the next pressure, and so on. This generates a \filename{clausius-predictions.csv} file with the pressure, converged melting temperature and number of heating iterations as the columns.

\begin{algorithm}[H]
\caption{Method to numerically integrate Clausius-Clapeyron equation.\label{algorithm:CC}}
\begin{algorithmic} [1]
\State Set temperature convergence threshold 
\For 
{$pressures=P^{0},P^{1}, \ldots, P^{N}$}
    \State Create solid, liquid configurations at converged/known coexistence point $(P^{i},T^{i})$
    \State Calculate $dP/dT$ from $\Delta H$ and $\Delta V$
    \State Use $dP/dT$ to make a straight-line approximation for next coexistence point $(P^{i+1},T_{g})$
    \State Run compression $(P^{i},T^{i}) \rightarrow (P^{i+1},T^{i})$, calculate $\Delta G_{C} $
    \State Set initial high $\Delta T$
    \While 
    {$\Delta T > threshold$}
        \State Run heating sim $(P^{i+1},T^{i}) \rightarrow (P^{i+1},T_{g})$, calculate $\Delta G_{H} $
        \State Use $\Delta G_{C} + \Delta G_{H}$ to obtain correction factor $\Delta T$
        \State Refine guess $T_{g} = T_{g} + \Delta T$
    \EndWhile
    \State Predict converged melting temperature $T^{i+1} = T_{g}$ at $P^{i+1}$
\EndFor
\end{algorithmic}
\end{algorithm}

\section*{References}
\bibliographystyle{apsrev4-2}
\bibliography{main}